\documentclass[aip,jap,reprint,longbibliography]{revtex4-1}

\usepackage{graphicx}
\usepackage{amsmath,amssymb,bm}
\usepackage{mathtools}
\usepackage[separate-uncertainty]{siunitx}

\DeclareMathAlphabet{\mathup}{OT1}{\familydefault}{m}{n} 
\newcommand*{\mup}[1]{\mathup{#1}} 
\newcommand*{\diff}{\,\mathup{d}}
\newcommand*{\iu}{\mathup{i}\mkern1mu}

\DeclareMathOperator{\e}{e}

\newcommand\ten[1]{\bm{\mathrm{#1}}}
\newcommand\dirvec[1]{\mathbf{e}_{#1}}

\usepackage{xcolor}
\usepackage[hidelinks]{hyperref}
\usepackage[version=4]{mhchem}
\begin{document}

\title{Electroelastic guided wave dispersion in piezoelectric plates: spectral methods and laser-ultrasound experiments}

\author{D. A. Kiefer}
\affiliation{Institut Langevin, ESPCI Paris, Université PSL, CNRS, 75005 Paris, France}
\author{G. Watzl}
\affiliation{Research Center for Non Destructive Testing GmbH, Altenberger Str. 69, 4040 Linz, Austria}
\author{K. Burgholzer}
\affiliation{
Institute of Semiconductor and Solid State Physics, Johannes Kepler University Linz, Altenberger Str. 69, 4040 Linz, Austria
}
\author{M. Ryzy}
\affiliation{Research Center for Non Destructive Testing GmbH, Altenberger Str. 69, 4040 Linz, Austria}
\author{C. Grünsteidl}
\email{clemens.gruensteidl@recendt.at}
\affiliation{Research Center for Non Destructive Testing GmbH, Altenberger Str. 69, 4040 Linz, Austria}

\date{\today}

\begin{abstract}
Electroelastic waves in piezoelectric media are widely used in sensing and filtering applications. 
Despite extensive research, computing the guided wave dispersion remains challenging.
This paper presents semi-analytical approaches based on spectral methods to efficiently and reliably compute dispersion curves. 
We systematically assess the impact of electrical boundary conditions on a 128° Y-cut \ce{LiNbO3} wafer, examining open-open, open-shorted and shorted-shorted surfaces configurations. 
Multi-modal dispersion maps obtained from laser-ultrasonic experiments for each boundary condition exhibit excellent agreement with the computational predictions. 
A straightforward implementation of the spectral collocation method is made available as \texttt{GEW piezo plate} (\url{https://doi.org/10.5281/zenodo.14205789}), while the spectral element method will be integrated to \texttt{GEWtool} (\url{http://doi.org/10.5281/zenodo.10114243}). 
Therewith, we aim to make advanced semi-analytical techniques more accessible to physicists and engineers relying on dispersion analysis.
\end{abstract}

\pacs{}

\maketitle

\section{\label{sec:intro}Introduction}
In a waveguide, waves are constrained to travel along a defined path due to boundaries that channel the energy~\cite{Auld2}. Guided elastic waves have applications in non-destructive testing~\cite{wilcoxModeTransducerSelection2001}, material characterization~\cite{Clorennec2007,Thelen2021,Grabec2024,Watzl2025} and sensing~\cite{ceglaMaterialPropertyMeasurement2005,kieferTransitTimeLamb2022}.
If the waveguide's material is piezoelectric, the mechanical and electrical fields are coupled, resulting in electroelastic waves~\cite{Auld1,Auld2,kuangElectroelasticWave2014,kyameWavePropagationPiezoelectric1949,epsteinElasticWaveFormulationElectroelastic1973}.
These have found widespread application in micro-electromechanical systems (MEMS), e.g., as surface acoustic wave (SAW) devices for signal processing and sensing~\cite{mandalSurfaceAcousticWave2022}. In other devices, like bulk acoustic wave filters, spurious guided modes need to be avoided~\cite{Kokkonen2010}.
Due to technological advances in materials, thin film technology and further development of MEMS devices, application-driven research on 
guided waves in piezoelectric plates is still very active today~\cite{Yao2024,luMicromachinedPiezoelectricLamb2023, Setiawan2021,Caliendo2017}.

Electroelastic waves in piezoelectric media are well studied. The theoretical fundamentals are covered in classic textbooks \cite{Auld1,Auld2}. Initial work describes the propagation of bulk waves~\cite{kyameWavePropagationPiezoelectric1949,epsteinElasticWaveFormulationElectroelastic1973}. Work on modeling guided waves in piezoelectric plates was pioneered by Tiersten~\cite{tierstenWavePropagationInfinite1963}, who considered metallized surfaces while restricting propagation to two specific directions. Later, fundamental modes propagating along principal directions with open (unmetallized) and shorted (metallized) surfaces were studied by Joshi et al.~\cite{Joshi1991} and validated experimentally with interdigital transducers. Syngellakis and Lee~\cite{syngellakisPiezoelectricWaveDispersion1993} extended the modeling to arbitrary anisotropy and propagation direction and computed the dispersion of electrically shorted higher-order modes. Electrically open modes at different propagation directions were computed by Kuznetsova et al. \cite{Kuznetsova2004}.
Sun et al. \cite{Sun1998} presented solutions for open-open, open-shorted and shorted-shorted surfaces. A periodic grating of open and shorted regions can be used to design phononic crystals with tunable band gaps~\cite{vasseurElectricalEvidenceTunable2018}. Guo et al.~\cite{guoGuidedWavePropagation2009} analyzed multilayered piezoelectric plates based on a state-space formalism. Lastly, Zhu et al.~\cite{zhuInfluenceSurfaceConductivity2020} accounted for the electrical losses in the metallized surfaces of plates.

Root-searching of the characteristic equations was used in all previously mentioned studies (except the phononic crystal) to compute dispersion curves. Although this method is known to be slow and to miss solutions, it is still prevalent today. A software based on root-searching was made available in 1990 by Adler~\cite{adlerPCSoftwareSAW1990} and is still actively being used~\cite{caliendoUVSensorsBased2023,ageikinAnalysisRadiationAbsorption2023}. An alternative treatment of the transcendental characteristic equations are analytic approximations for limiting cases\cite{shuvalovLowfrequencyDispersionFundamental2010}.

In contrast to root-searching of the characteristic equations, semi-analytical methods discretize the waveguide cross section and then solve a standard algebraic eigenvalue problem to reliably and efficiently obtain solutions. Various discretization methods have been used for semi-analytical models of piezoelectric waveguides. Finite Elements were used very early by Lagasse~\cite{lagasseFiniteElementAnalysis1973} for ridge and wedge guides, while Cortes et al.~\cite{Cortes2008} treated multi-layered plates. Spectral methods are based on a global function basis to expand the field and are particularly well-suited for layered structures. In this context, Laguerre polynomials were used for surface waves~\cite{dattaAnalysisSurfaceWaves1978,lefebvreConceptualAdvantagesLimitations1998} and Legendre polynomials~\cite{zhangFullDispersionCharacteristics2019} for plates. Laude et al.~\cite{laudeComputationPlateWave2010} used a Fourier series representation but encountered difficulties due to the non-vanishing electrical field outside the plate. Despite the advantages of semi-analytical methods, their wide-spread adoption is hindered by the lack of publicly available implementations and the relatively high level of knowledge on numerical computing required to implement them.

From an experimental point of view, laser-based ultrasound (LUS) has proven to be a valuable tool to precisely resolve the multi-modal dispersion curves of plates\cite{Thelen2021, Pierce1997, Schoepfer2013, Rohringer2019, Claes2021} and to access specific modes in the response spectrum \cite{Prada2005, Clorennec2007, Watzl2022a, Ryzy2023, Watzl2025, kieferBeatingResonancePatterns2023a}. 
Importantly, the contact-free operation of LUS allows for flexible scanning of the wave field under various environmental and boundary conditions. This is particularly valuable for model validation and for inverse characterization of material parameters.
We are aware of two studies~\cite{Yang2006,chenLambWaveDispersion2011} that present LUS measurements of piezoelectric plates, but the effect of electrical BCs was not examined therein.

In this work, we present a semi-analytical model for guided electroelastic waves in piezoelectric plates based on spectral methods. Our focus lies on the highly instructive spectral collocation method (SCM)~\cite{trefethenSpectralMethodsMATLAB2000,fornbergPracticalGuidePseudospectral1996,weidemanMATLABDifferentiationMatrix2000}. 
The SCM was previously shown to be efficient and robust to compute dispersion of purely elastic waves~\cite{adamouSpectralMethodsModelling2004,hernandoquintanillaModelingGuidedElastic2015}. 
Its main advantage is the straight-forward implementation based on the underlying partial differential equations in their conventional strong form.
The SCM code by Kiefer~\cite{GEWdispersionscript} is extended to piezoelectric plates and we make our implementation publicly available under the name \texttt{GEW piezo plate}~\cite{GEWpiezoplate}. 
As an alternative, highly related method we also introduce the spectral element method (SEM)~\cite{fichtnerFullSeismicWaveform2010,gravenkampNumericalApproachComputation2012} in Appendix~\ref{sec:spectral_elements}. While it is numerically superior to the SCM, it is at the same time conceptually more intricate. Our SEM implementation will be provided in \texttt{GEWtool}~\cite{kieferGEWtool2023}, a software for guided wave analysis. 
Overall, we hope to demonstrate with this contribution the power of spectral methods, the simplicity of SCM, and contribute to a wider adoption of semi-analytical methods for dispersion computations. 

Furthermore, we validate our model with LUS experiments and contribute a compilation of multi-modal guided wave spectra in \ce{LiNbO3} wafers, including two different propagation directions and three different electrical BCs.

The paper is organized in the following way:
in Sec.~\ref{sec:modeling} we derive the guided wave problem in piezoelectric media, with focus on the treatment of electrical BCs.
The implementation of the SCM

is explained in Sec.~\ref{sec:scm} and computational results are presented in Sec.~\ref{sec:results_wafer}.

Our LUS system is described in Sec.~\ref{sec:experiments}. Lastly, measurement results are presented and discussed in Sec.~\ref{sec:results} before we conclude in Sec.~\ref{sec:conclusion}.

\section{\label{sec:modeling} Modeling electroelastic waves}
Assuming a quasi-static electric field, we model waves in the piezoelectric medium in terms of the mechanical displacements~$\ten{\tilde{u}}$ and the electric potential~$\tilde{\phi}$. The electric field intensity is $-\nabla\tilde{\phi}$. The piezoelectric constitutive equations relate the mechanical stress~$\ten{\tilde{T}}$ and electric flux density~$\ten{\tilde{D}}$ to $\ten{\tilde{u}}$ and $\tilde{\phi}$ by~\cite{Auld1,tierstenWavePropagationInfinite1963,rupitschPiezoelectricSensorsActuators2018}
\begin{subequations}\label{eq:piezo_constitutive}
\begin{align}
    \label{eq:stress_piezo1}
    \ten{\tilde{T}} &= \ten{c}:\nabla \ten{\tilde{u}} + \nabla \tilde{\phi} \cdot \ten{e} \,, \\
    \label{eq:elecFlux_piezo2}
    \ten{\tilde{D}} &= \ten{e}:\nabla\ten{\tilde{u}} - \ten{\epsilon} \cdot \nabla \tilde{\phi} \,,
\end{align}
\end{subequations}
where $\ten{c}$ is the 4th-order stiffness tensor at constant electric field intensity, $\ten{e}$ the 3rd order piezoelectric coupling tensor and $\ten{\epsilon}$ the 2nd-order permittivity tensor at constant mechanical strain~\cite{rupitschPiezoelectricSensorsActuators2018}.

For a time-harmonic field at angular frequency~$\omega$, the balance of linear momentum, i.e, $\nabla \cdot \ten{\tilde{T}} + \rho \omega^2 \ten{\tilde{u}} = 0$, must hold given the mass density~$\rho$. Moreover, without free electrical charges in the material $\nabla \cdot \ten{\tilde{D}} = 0$. In account of (\ref{eq:piezo_constitutive}), the governing equations are~\cite{Auld2,tierstenWavePropagationInfinite1963}
\begin{subequations}\label{eq:governing}
\begin{align}
    \nabla \cdot (\ten{c}:\nabla \ten{\tilde{u}}) + \nabla \cdot (\nabla \tilde{\phi} \cdot \ten{e}) + \rho \omega^2 \ten{\tilde{u}} &= \ten{0} \,, \\
    \nabla \cdot (\ten{e}:\nabla\ten{\tilde{u}}) - \nabla \cdot (\ten{\epsilon} \cdot \nabla \tilde{\phi}) &= 0 \,.
\end{align}
\end{subequations}

\begin{figure}[tb]
    \centering
    \includegraphics[width=0.8\linewidth]{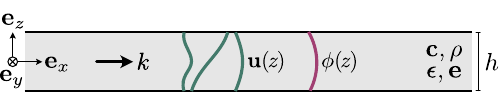}\\
    (a) electrically shorted plate (metallized)\\[1em]
    \includegraphics[width=0.8\linewidth]{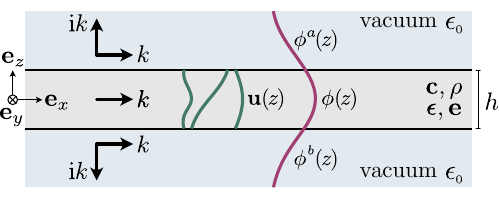}\\
    (b) electrically open plate (non-metallized)
    
    \caption{Cross-sectional sketch of the piezoelectric plate. (a)~Electrically shorted surfaces confine the field inside the plate, while (b)~non-metallized surfaces lead to a nonzero electric field in the surrounding vacuum.}
    \label{fig:plate_sketch}
\end{figure}

We are interested in plane waves with wavenumber~$k$ propagating in a plate of thickness~$h$ but otherwise infinite extent, see Fig.~\ref{fig:plate_sketch}. The $\dirvec{x}\dirvec{y}\dirvec{z}$-coordinate system (where $\dirvec{i}$ with one index are the unit directional vectors) is aligned with the wave propagation such that the wave vector is along $\dirvec{x}$ and $\dirvec{z}$ is the normal to the plate's surfaces. The spatio-temporal field is then of the form
\begin{subequations}\label{eq:ansatz}
\begin{align}
    \ten{\tilde{u}}(x,y,z,t) &= \ten{u}(k,z,\omega)\e^{\iu k x - \iu \omega t} \\
    \tilde{\phi}(x,y,z,t) &= \phi(k,z,\omega)\e^{\iu k x - \iu \omega t} \,,
\end{align}
\end{subequations}
with $z \in (-h/2, h/2)$.
We proceed as was done for the elastic case in Ref.~\cite{kieferComputingZerogroupvelocityPoints2023}. 
In account of the ansatz (\ref{eq:ansatz}), we can use $\nabla \coloneq \iu k \dirvec{x} + \partial_z \dirvec{z}$ in (\ref{eq:governing}). Multiplying out, defining the 2nd-order tensors $\ten{c}_{ij} = {\dirvec{i} \cdot \ten{c} \cdot \dirvec{j}}$, the 1st-order tensors $\ten{e}_{ij} = {\dirvec{i} \cdot \ten{e} \cdot \dirvec{j}}$ and the scalars $\epsilon_{ij} = {\dirvec{i} \cdot \ten{\epsilon} \cdot \dirvec{j}}$ and re-grouping we obtain for the\footnote{Note that we exploited the symmetry $e_{ijk} = e_{ikj}$ in the definition of $\ten{e}_{ij} := \dirvec{j}\cdot(\dirvec{i}\cdot\ten{e}) = \dirvec{i} \cdot \ten{e} \cdot \dirvec{j}$.}
\begin{subequations}\label{eq:governing_waveguide}
\begin{align}
    &\text{balance of linear momentum:} \nonumber \\
    \label{eq:governing_waveguide1}
    &[(\iu k)^2 \ten{c}_{xx} + \iu k (\ten{c}_{xz} + \ten{c}_{zx}) \partial_z + \ten{c}_{zz} \partial_z^2 ] \cdot \ten{u}  + \rho \omega^2 \ten{u} \nonumber \\ 
    & \quad+ [(\iu k)^2 \ten{e}_{xx} + \iu k (\ten{e}_{xz} + \ten{e}_{zx}) \partial_z + \ten{e}_{zz} \partial_z^2 ] \phi = \ten{0} \,, \\
    \label{eq:governing_waveguide2}
    &\text{balance of charges:} \nonumber \\
    &[(\iu k)^2 \ten{e}_{xx} + \iu k (\ten{e}_{xz} + \ten{e}_{zx}) \partial_z + \ten{e}_{zz} \partial_z^2 ] \cdot \ten{u} \nonumber \\
    & \quad + [(\iu k)^2 \ten{\epsilon}_{xx} + \iu k (\ten{\epsilon}_{xz} + \ten{\epsilon}_{zx}) \partial_z + \ten{\epsilon}_{zz} \partial_z^2] \phi = 0 \,.
\end{align}
\end{subequations}
The above equations must hold across the thickness of the plate, i.e., for $z \in (-h/2, h/2)$. Note that common materials are non-gyrotropic~\cite{weisLithiumNiobateSummary1985}, meaning that their off-diagonal permittivity components $\epsilon_{xz}$ and $\epsilon_{zx}$ vanish. Nevertheless, we retain the corresponding term in (\ref{eq:governing_waveguide2}) for the sake of generality.

BCs are needed to complement (\ref{eq:governing_waveguide}). The plate shall be mechanically free, i.e., the tractions $\dirvec{z} \cdot \ten{\tilde{T}}$ vanish at the surfaces $z = \pm h/2$. Using (\ref{eq:stress_piezo1}) and the plane wave ansatz (\ref{eq:ansatz}), this reads

\begin{equation}\label{eq:traction_free}
    \iu k [\ten{c}_{zx} \cdot \ten{u} + \ten{e}_{xz} \phi ]  + \ten{c}_{zz} \partial_z \cdot \ten{u}  + \ten{e}_{zz} \partial_z \phi = \ten{0},\ z = \pm h/2 \,.
\end{equation}
From an electrical point of view, two different kind of BCs will be studied: (i)~shorted/metallized surfaces that model an infinitely thin, perfectly conducting coating as sketched in Fig.~\ref{fig:plate_sketch}a and (ii)~the electrically open (non-metallized) surface as shown in Fig.~\ref{fig:plate_sketch}b. Any combination of these two conditions is possible at the two surfaces of the plate. If the plate is shorted at $z = \pm h/2$,
\begin{equation}\label{eq:potentialZero_Dirichlet}
    \phi = 0 \text{ at } z = \pm h/2 \,.
\end{equation}
The open case is somewhat more intricate. Often, $\dirvec{z} \cdot \ten{\tilde{D}} = 0$ has been used~\cite{dattaAnalysisSurfaceWaves1978,lefebvreConceptualAdvantagesLimitations1998}, forcing the electric energy to be confined within the plate. Except for $k h \ll 1$, this tends to be a very good approximation because the permittivity of the piezoelectric material is usually much higher than that in vacuum~\cite{lagasseFiniteElementAnalysis1973}. But the electric field in the exterior vacuum is actually nonzero and this should be accounted for with appropriate interface conditions at the plate's surfaces. As has been shown previously~\cite{joshiPropagationUltrasonicLamb1991,zhuInfluenceSurfaceConductivity2020}, this amounts to modeling nonhomogeneous Neumann BCs involving the wavenumber. We pursue this approach, as the computational costs are similar and the modeling not very complicated.

As the field is a plane harmonic wave in the plate, this must also be the case in the exterior. For $z > h/2$ we have $\tilde{\phi}^{a} = \phi^{a} \e^ {\iu k_z^{a} (z-h/2) + \iu k x - \iu \omega t}$, where Snell's law has been used. In the vacuum, $\ten{\tilde{D}}^{a} = -\epsilon_0 \nabla \tilde{\phi}^{a}$ and $\nabla \cdot \ten{\tilde{D}}^{a} = 0$, from where $k_z^{a} = \pm \iu k$ is obtained. $\epsilon_0$ is the permittivity of vaccum. The sign of $k_z^{a}$ is chosen such that $\phi^{a} \to 0$ as $z \to \infty$. A similar analysis for the field below the plate shows that the opposite sign must be chosen for $k_z^{b}$. Note that the field is purely evanescent along $z$ and carries no energy away from the plate, in accordance with the quasi-static approximation. If the plate is open at the top surface, the normal electrical flux density is required to be continuous across this interface, i.e., $\dirvec{z} \cdot \ten{\tilde{D}} = \dirvec{z} \cdot \ten{\tilde{D}}^{a/b}$ at $h/2$. Using (\ref{eq:elecFlux_piezo2}) and (\ref{eq:ansatz}), the continuity at $z = h/2$ reads
\begin{equation}
    \dirvec{z}\cdot\ten{D} = [\iu k \ten{e}_{zx} + \ten{e}_{zz} \partial_z] \cdot \ten{u} - [\iu k \epsilon_{zx} + \epsilon_{zz} \partial_z] \phi = -\iu \epsilon_0 \iu k \phi^{a} \,,
\end{equation}
which represents an \emph{inhomogeneous} Neumann BC (the right-hand side is nonzero).
Next, we also impose continuity of the electrical potential, i.e., $\phi^{a} = \phi(h/2)$. Accounting for this and re-arranging in terms of $\iu k$ finally yields the open-plate conditions at $z = \pm h/2$, namely,
\begin{subequations}\label{eq:continuity_elecFlux}
\begin{align}
    \iu k [\ten{e}_{zx} \cdot \ten{u} - \epsilon_{zx} \phi \pm \iu \epsilon_0 \phi] + \ten{e}_{zz} \partial_z \cdot \ten{u}  - \epsilon_{zz} \partial_z \phi &= 0 \,,
\end{align}
\end{subequations}
where the positive sign is for the top surface at $z > 0$. Note that with $\epsilon_0 = 0$, this condition is identical to the commonly used approximation $\dirvec{z} \cdot \ten{\tilde{D}} = 0$ at the surface.

\section{\label{sec:scm}Calculation of dispersion curves using the SCM}
The computational method proceeds in three steps: 
\begin{enumerate}
    \item Re-write the governing equations (\ref{eq:governingSystem}) and the Neumann BCs (\ref{eq:NeumannBCSystem}) in matrix form.
    \item Discretize both differential systems.
    \item Incorporate the discrete BCs into the governing system. Optionally, replace the Neumann BC with Dirichlet BCs.
\end{enumerate}
In this way, an algebraic linear system is obtained that can be solved on a computer using standard techniques in order to obtain the desired guided wave solutions. 

First, we collect the unknown field variables in the ${4 \times 1}$-block matrix (first block of size 3, second block of size 1)
\begin{equation}
    \Psi = \begin{bmatrix}
        \ten{u} \\
        \phi
    \end{bmatrix} \,,
\end{equation}
and assemble the governing equations (\ref{eq:governing_waveguide1}) and (\ref{eq:governing_waveguide2}) into one matrix equation in terms of $\Psi$, which yields
\begin{align}\label{eq:governingSystem}
    \left[(\iu k)^2 \mathcal{A} + \iu k (\mathcal{B} + \mathcal{B}^T) \partial_z + \mathcal{C} \partial_z^2 + \omega^2 \mathcal{M} \right] \Psi &= 0 \,,
\end{align}
with matrices
\begin{subequations}\label{eq:governingMats}
\begin{align}
    \mathcal{A} &= \begin{bmatrix}
        \ten{c}_{xx}    &  \ten{e}_{xx} \\
        \ten{e}_{xx}^T  &  -\epsilon_{xx}
    \end{bmatrix} \,, &
    \mathcal{B} &= \begin{bmatrix}
        \ten{c}_{zx}    &  \ten{e}_{xz} \\
        \ten{e}_{zx}^T  &  -\epsilon_{zx}
    \end{bmatrix}\\
    \mathcal{C} &= \begin{bmatrix}
        \ten{c}_{zz}    &  \ten{e}_{zz} \\
        \ten{e}_{zz}^T  &  -\epsilon_{zz}
    \end{bmatrix} \,, &
    \mathcal{M} &= \begin{bmatrix}
        \rho\ten{I}    &  0 \\
        0  &  0
    \end{bmatrix} \,.
\end{align}
\end{subequations}
Thereby, we have exploited the symmetries $\ten{c}_{xz} = \ten{c}_{zx}^T$ and $\epsilon_{xz} = \epsilon_{zx}$ in writing the term in $\mathcal{B} + \mathcal{B}^T$ (but note that $\ten{e}_{xz} \ne \ten{e}_{zx}^T$ in $\mathcal{B}$). Moreover, $\ten{I}$ is the 2nd-order identity tensor, ``$0$'' are zero-matrices of appropriate size and all tensors are to be interpreted as the corresponding coefficient matrices. Assembling in a similar way the mechanical and electrical Neumann BCs (\ref{eq:traction_free}) and (\ref{eq:continuity_elecFlux}) into a matrix equation gives
\begin{equation}\label{eq:NeumannBCSystem}
    \left[ \iu k (\mathcal{B} + \mathcal{V}) + \mathcal{C} \partial_z \right] \Psi = 0 \,,
\end{equation}
where $\mathcal{V}$ is the block matrix 
\begin{align}\label{eq:NeumannBCMats}
    \mathcal{V} &= \begin{bmatrix}
        0   &       0                   \\
        0   &  \pm \iu \epsilon_0
    \end{bmatrix} \,,
\end{align}
which accounts for the nonzero electrical flux density into the vacuum domain.
The positive/negative sign is for the top/bottom boundary respectively. For the moment, we do not need to deal with the Dirichlet BC (\ref{eq:potentialZero_Dirichlet}) that describes shorted electrodes as it will be implemented after discretization by eliminating the corresponding degree of freedom, i.e., $\phi(\pm h/2)$, which is known in advance to be zero. 

Second, a large number of numerical discretization techniques could be used to obtain algebraic approximations to the boundary value problem (\ref{eq:governingSystem}) and (\ref{eq:NeumannBCSystem}). For the numerically superior SEM we refer to Appendix~\ref{sec:spectral_elements} and instead focus on the more instructive SCM in the following. The SCM is particularly simple to implement as it does not require a weak formulation nor previous meshing of the geometry. Furthermore, it exhibits exponential convergence when increasing the degrees of freedom, leading to small matrices. 
The Chebyshev SCM approximates an unknown function $\phi(z)$ by a weighted superposition of the first $N$ Chebyshev polynomials and the error is then minimized on $N$ so-called Chebyshev-Gauß-Lobatto collocation points $z_i$, $i \in \{1...N\}$. For details on the SCM refer to the standard literature~\cite{trefethenSpectralMethodsMATLAB2000,fornbergPracticalGuidePseudospectral1996}. For implementation purposes, it is important that the method allows for the explicit construction of a differentiation matrix $D_z$. To explain its meaning, assume that $\phi_{\mathup{d}} = [\phi(z_i)]$ is the mathematical vector that collects the values of $\phi(z)$ at all collocation points $z_i$; and $\phi'_{\mathup{d}}$ is the vector collecting the corresponding values of $\partial_z \phi(z)$. The differentiation matrix relates these two vectors through $\phi'_{\mathup{d}} \approx D_z \phi_{\mathup{d}}$, providing a discrete counterpart to the differentiation operation. We use the Matlab package \texttt{DMSUITE}~\cite{weidemanMATLABDifferentiationMatrix2000} by Weideman and Reddy to generate the differentiation matrices $D_z$ and $D_z^2$ of size $N \times N$.

The discretization of (\ref{eq:governingSystem}) is performed by formally mapping $\partial_z \mapsto D_z$, $\partial_z^2 \mapsto D_z^2$ and for the $\partial_z$-independent terms $1 \mapsto I_{\mathup{d}}$, where $I_{\mathup{d}}$ is the $N \times N$-identity matrix. Thereby, the component-wise multiplication of $\partial_z$, $\partial_z^2$ or $1$ with the matrices $\mathcal{A}$, $\mathcal{B}$, $\mathcal{C}$ or $\mathcal{M}$ become Kronecker products
\footnote{The Kronecker product $A\otimes B$ of the $n \times m$ matrix $A$ and the $p \times q$ matrix $B$ is defined as the matrix of size $np\times mq$ given in block form by
\begin{equation}
    A\otimes B = \left[\begin{matrix}
    A_{11}B & \cdots & A_{1m}B\cr
    \vdots & & \vdots\cr
    A_{n1}B & \cdots & A_{nm}B
\end{matrix}\right].
\end{equation}
}, which we denote by ``$\otimes$''.
Overall, the discrete approximation to (\ref{eq:governingSystem}) is
\begin{align}\label{eq:governingDiscrete}
    \left[(\iu k)^2 \tilde{L}_2 + \iu k \tilde{L}_1 + \tilde{L}_0 + \omega^2 \tilde{M} \right] \Psi_{\mathup{d}} &= 0 \,,
\end{align}
with $4N \times 4N$ matrices
\begin{align}
    \tilde{L}_2 &= \mathcal{A} \otimes I_{\mathup{d}} \,, & 
    \tilde{L}_1 &= (\mathcal{B} + \mathcal{B}^T) \otimes D_z \,, \\
    \tilde{L}_0 &= \mathcal{C} \otimes D_z^2 \,, &
    \tilde{M}   &= \mathcal{M}   \otimes I_{\mathup{d}} \,.
\end{align}
Proceeding in the same way for the Neumann BC in (\ref{eq:NeumannBCSystem}) yields
\begin{equation}\label{eq:NeumannBCDiscrete}
    \left[ \iu k B_1 + B_0 \right] \Psi_{\mathup{d}} = 0 \,,
\end{equation}
with 
\begin{align}
    B_1 &= (\mathcal{B} + \mathcal{V}) \otimes I_{\mathup{d}} \,, & 
    B_0 &= \mathcal{C} \otimes D_z \,. 
\end{align}

Third, only the equations at $z = \pm h/2$ are needed from (\ref{eq:NeumannBCDiscrete}), i.e., the equation numbers $j \in \{1, N, N+1, 2 N, 2 N +1, 3 N, 3 N +1, 4 N\}$.
These BCs are incorporated into the discrete algebraic system (\ref{eq:governingDiscrete}) by replacing the rows $j$ of $\tilde{L}_i$ and $\tilde{M}$  with the corresponding rows of $B_i$ or zero if the term is not present in (\ref{eq:NeumannBCDiscrete}).  The resulting matrices are denoted $L_2$, $L_1$, $L_0$ and $M$. In this way, we obtain the linear system 
\begin{align}\label{eq:waveguideProblemDiscrete}
    \left[(\iu k)^2 L_2 + \iu k L_1 + L_0 + \omega^2 M \right] \Psi_{\mathup{d}} &= 0 \,
\end{align}
that fully describes plane harmonic electroelastic waves in the plate. For now, the surfaces are both electrically open as depicted in Fig.~\ref{fig:plate_sketch}b. If the plate is to be shorted at the surface corresponding to the \hbox{$i$th} collocation point $z_i$, we simply remove the \hbox{${3 N + i}$-th} row and column of all matrices (the first $3N$ correspond to displacement degrees of freedom), thereby replacing the open condition with the Dirichlet condition given in (\ref{eq:potentialZero_Dirichlet}). 

Equation (\ref{eq:waveguideProblemDiscrete}) represents an algebraic eigenvalue problem. By fixing $k = k_0$ we may compute the eigenvalues $\omega^2_n$ with standard techniques, e.g., \texttt{eig} in Matlab. We remark that $M$ is a singular matrix, leading to some restrictions on the choice of the eigenvalue solver. Nonetheless, modern methods appropriately handle this case. It is also possible to do the reverse and fix the angular frequency $\omega = \omega_0$ in order to compute the eigenvalues $\iu k_n$. This quadratic eigenvalue problem can be solved by standard companion linearization techniques as implemented in Matlab's \texttt{polyeig}. Note that even for real frequencies this approach yields a \emph{complex-valued} wavenumber spectrum~\cite{Auld2}. In either case, the eigenvectors $\Psi_{\mathup{d}}$ provide access to the mechanical and electrical field distributions $\ten{u}(z_i)$ and $\phi(z_i)$. 

Our computations show very good agreement with those published by Kuznetsova et al.\cite{Kuznetsova2004}, see Appendix\,\ref{sec:verification}. Note that the higher the modal order (i.e., the higher the frequency), the larger the number of discretization points $N$ needs to be. To ensure 6-digits accuracy with the SCM, we use $N = 20$ in all subsequent computations. A convergence analysis is also presented in Appendix\,\ref{sec:verification}.

\begin{figure*}[tb]
    \centering
    \includegraphics[width=1\linewidth]{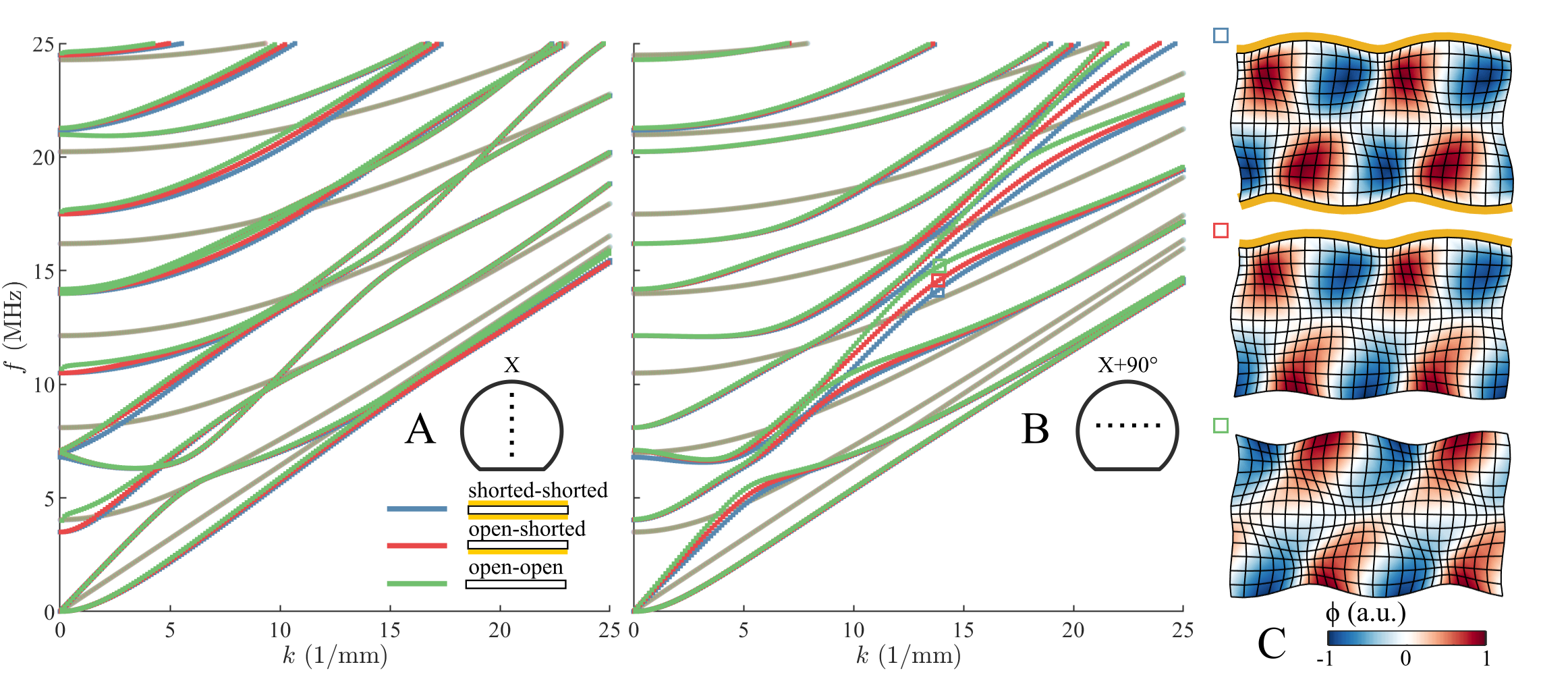}
    \caption{Dispersion curves of electroelastic waves propagating in a 128° Y-cut \ce{LiNbO3} plate. Three different sets of electrical BCs are compared: shorted-shorted, open-shorted, open-open. The SH-like waves are grayed-out. A: propagation along the $X$-axis, B: propagation orthogonal to the $X$-axis, C: Displacement mode shapes (grid) and electric potential (colormap) for the points marked in B.}
    \label{fig:dispersion_compare_bc}
\end{figure*}

\section{\label{sec:results_wafer}Waves in a \texorpdfstring{\ce{LiNbO3}}{LiNbO3} wafer}
We now investigate guided waves in a $\qty{510}{\micro\meter}$-thick monocrystalline 128° Y-cut lithium niobate (\ce{LiNbO3}) wafer. This material is often used for MEMS as it exhibits large electromechanical coupling~\cite{weisLithiumNiobateSummary1985,shibayamaOptimumCutRotated1976}. The parameters are given in Appendix~\ref{sec:material}.
To account for the crystal orientation in the computation, we need to rotate the tensors $\ten{c}$, $\ten{e}$, $\ten{\epsilon}$--which are initially given~\cite{kovacs1990} in the orthonormal $\dirvec{X}\dirvec{Y}\dirvec{Z}$-material coordinate system (\autoref{fig:setupSample}) to the wave propagation frame $\dirvec{x}\dirvec{y}\dirvec{z}$ (Fig.~\ref{fig:plate_sketch}). Here, $\dirvec{X}$ lies within the wafer surface and is perpendicular to the wafer flat. The material's $\dirvec{Z}$-axis ($\dirvec{Y}$-axis) has been rotated out of the plate's normal $\dirvec{z}$ by an angle of $\mu = \text{38°}$ ($\mu = \text{128°}$) around $\dirvec{X}$.

Two directions of propagation are considered: (i) along $X$ and (ii) along $X+\qty{90}{\degree}$, i.e., along the wafer flat, see \autoref{fig:setupSample}. 
Furthermore, three different combinations of electrical BCs are systematically investigated: shorted-shorted, open-shorted and open-open. The different BCs are compared in Fig.~\ref{fig:dispersion_compare_bc}A for propagation along $X$ and in Fig.~\ref{fig:dispersion_compare_bc}B for propagation in $X+\qty{90}{\degree}$. Several regions of the dispersion curves are observed to be quite strongly affected by the BCs, especially for propagation in $X+\qty{90}{\degree}$. Strong electromechanical coupling can be expected in these regions. The SCM computation also provides the mechanical and electrical modal field distributions. Three examples picked at the marks in Fig.~\ref{fig:dispersion_compare_bc}B are depicted in Fig.~\ref{fig:dispersion_compare_bc}C. Thereby, the field was extruded in propagation direction according to the harmonic ansatz from (\ref{eq:ansatz}).

\section{\label{sec:experiments}Sample preparation and LUS system}
The samples investigated were SAW-grade \qty{128}{\degree} Y-cut \ce{LiNbO3} wafers with a nominal thickness of \qty{500}{\micro\meter}.
The shorted BCs were produced by depositing $\qty{3}{\nano\meter}$ titanium and $\qty{400}{\nano\meter}$ gold by physical vapor deposition carried out in a Balzers PLS 570, metallizing either one or both sides of the wafer.

\begin{figure}[tb]
    \centering
    \includegraphics[width=1\linewidth]{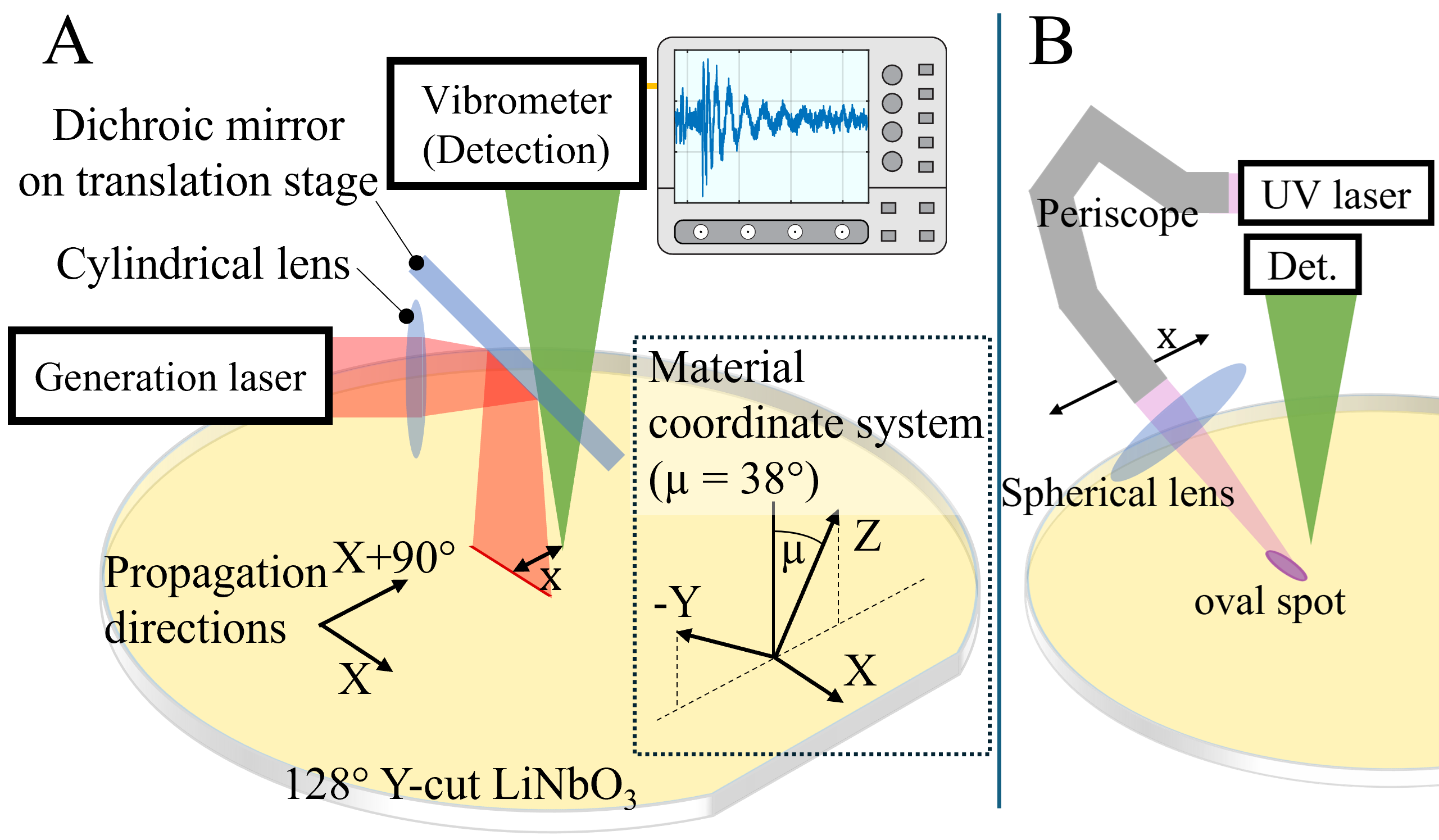}
    \caption{A: Investigated 128° Y-cut \ce{LiNbO3} sample and sketch of the LUS system. The orientation of the material coordinate system $(X,Y,Z)$ and the investigated directions of propagation ($X$ and $X+90$°) are shown. B: Workaround for uncoated sample using a UV laser at oblique incidence.}
    \label{fig:setupSample}
\end{figure}

We use a LUS system to record the spatio-temporal response of the plate. 
The LUS setup is sketched in \autoref{fig:setupSample}\,A.
To generate waves, we use a pulsed laser (Bright Solutions Wedge HB 1064) with a wavelength of $\qty{1064}{\nano\meter}$, pulse duration of about $\qty{1.5}{\nano\second}$, pulse energies of approximately $\qty{2}{\milli\joule}$, and a pulse repetition rate set to $\qty{1000}{\hertz}$.
The laser is focused by a cylindrical lens with $\qty{130}{\milli\meter}$ focal length, and deflected towards the sample by $\qty{90}{\degree}$ with a dichroic mirror which is mounted on an automated translation stage. By scanning the mirror, the distance between generation and detection is varied.
We deliberately slightly defocused the sample by about $\qty{3}{\milli\meter}$ with the intention to couple efficiently into modes of lower $k$ values and to avoid ablation.
The cylindrical lens produces a line shaped thermo-elastic source, predominantly emitting plane waves with a wave vector perpendicular to the line.

The out-of-plane component of the resulting surface displacement $u_z(x,t)$ is recorded with a two-wave-mixing vibrometer (Sound \& Bright Tempo-FS200) operating at $\qty{532}{\nano\meter}$ wavelength connected to an oscilloscope (Teledyne LeCroy WaveRunner HRO 66Zi).
The detection unit was focused onto the sample through the dichroic mirror with a lens of $\qty{100}{\milli\meter}$ focal length. 
A line of $\qty{14}{\milli\meter}$ was sampled along $x$ by scanning the dichroic mirror with a pitch of $\qty{50}{\micro\meter}$.
To improve the signal-to-noise ratio, 5000 traces were averaged for each measured position, and the bandwidth of the oscilloscope was limited to $\qty{50}{\mega\hertz}$.
\begin{figure}[tb]
    \centering
    \includegraphics[width=1\linewidth]{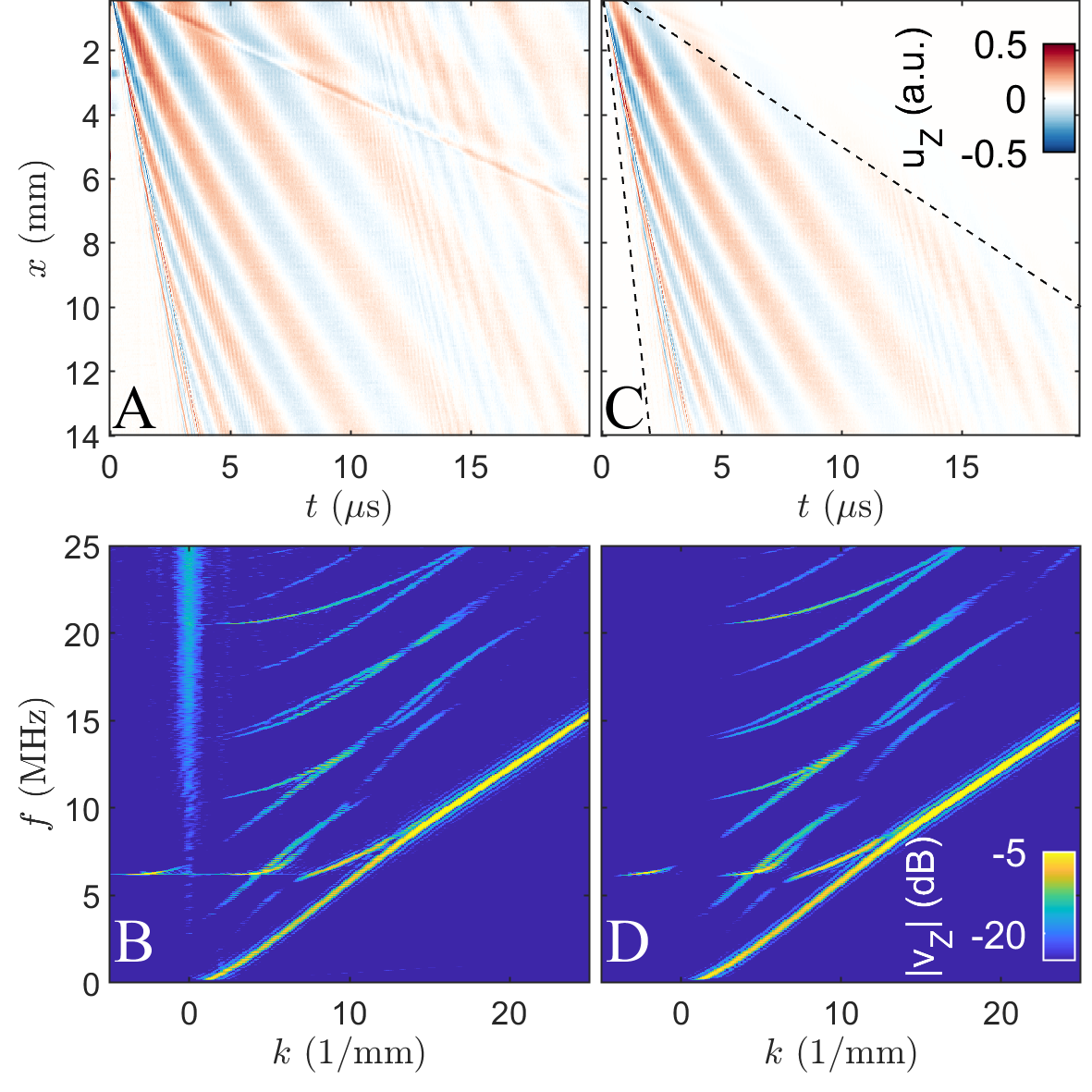}
    \caption{A: Recorded surface displacement $u_z(x,t)$ for the open-shorted sample, propagation in X-direction. B:~Frequency-wavenumber spectrum of A. C:~$u_z(x,t)$ from A, with windowing applied. Dashed lines indicate the selected signal range. D:~Spectrum of C, showing reduced noise and disturbances around $k=0$.}
    \label{fig:sigProc}
\end{figure}

\autoref{fig:sigProc}\,A shows the recorded $u_z(x,t)$ for the open-shorted case with propagation in X-direction. Applying a spatio-temporal Fourier transform results in a representation of the wavefield in the $k$-$f$-domain~\cite{Alleyne1991} and is shown in \autoref{fig:sigProc}\,B, where we plot the spectral velocity magnitude $|v_z(k,f)| = \omega |u_z(k,f)|$.

To enhance the experimental dispersion curves, we apply a window to the signals in the $x$-$t$-domain. 
The window is designed to select contributions with energy velocities between $\qty{500}{\meter/\second}$ and $\qty{7000}{\meter/\second}$ by cutting out a wedge of the data. To reduce side lobes, the window $w(x,t)$ is smoothed by convolution with a Hanning window in the $x$ and $t$ dimensions. \autoref{fig:sigProc}\,C shows the windowed signals $u_z(x,t) w(x,t)$ and also indicates the area selected by the limiting energy velocities as white dashed lines. The resulting dispersion maps with reduced noise and artefacts around $k=0$ are shown in \autoref{fig:sigProc}\,D. A portion of the dispersion spectrum around \qty{6}{\mega\hertz} has a negative wavenumber, but positive energy velocity and is thus detected. 
Such backward waves~\cite{kieferBeatingResonancePatterns2023a} appear in plates next to zero-group-velocity points. For simplicity, for \autoref{fig:LUS_SCM}  we symmetrize the measured spectra by $v_z(k,f) + v_z(-k,f)$ and only show positive values for $k$.

The open-open samples are almost transparent to the $\qty{1064}{\nano\meter}$ generation laser, resulting in insufficient coupling into guided waves. We followed two separate strategies to obtain measurements in this case: 
First, we coated the samples with black paint, which in some atepmts provided good results. Here we reduced the generation laser energy to avoid ablation (we otherwise observed thickness resonances as the predominant contributions in the spectrum). To enhance detection on the black painted sample, we added a small dot of silver paint, which we focused the vibrometer on.
Second, we used a $\qty{355}{\nano\meter}$ laser (Teem photonics PowerChip UV 355nm) in order to exploit the significantly higher absorption of \ce{LiNbO3} in the ultraviolet (UV) spectral range. Its pulse energy was $\qty{25}{\micro\joule}$, the pulse length about $\qty{1}{\nano\second}$ and the repetition rate was set to $\qty{250}{\hertz}$. The UV laser was guided to the sample by a flexible periscope with 7 mirrors. To scan the laser along the sample, the periscope's end-piece together with a $\qty{100}{\milli\meter}$ focal length spherical lens was mounted a linear scanning stage and pointed at the sample under an angle of $\approx$ $\qty{45}{\degree}$. This way, the primary setup described above could remain, while providing a workaround for the transparent sample. The workaround is shown in \autoref{fig:setupSample}\,B. 

With these two strategies we were able to produce dispersion maps of sufficient quality for comparison with the SCM. The best results are shown in \autoref{fig:LUS_SCM} (UV laser in C, black paint in F). We note that Yang et al.\cite{Yang2006} successfully measured this case using a $\qty{532}{nm}$ laser and a high pulse energy of $\qty{200}{mJ}$.

\section{\label{sec:results}Comparison of calculated and measured dispersion curves}
\begin{figure*}[tb]
    \centering
    \includegraphics[width=1\linewidth]{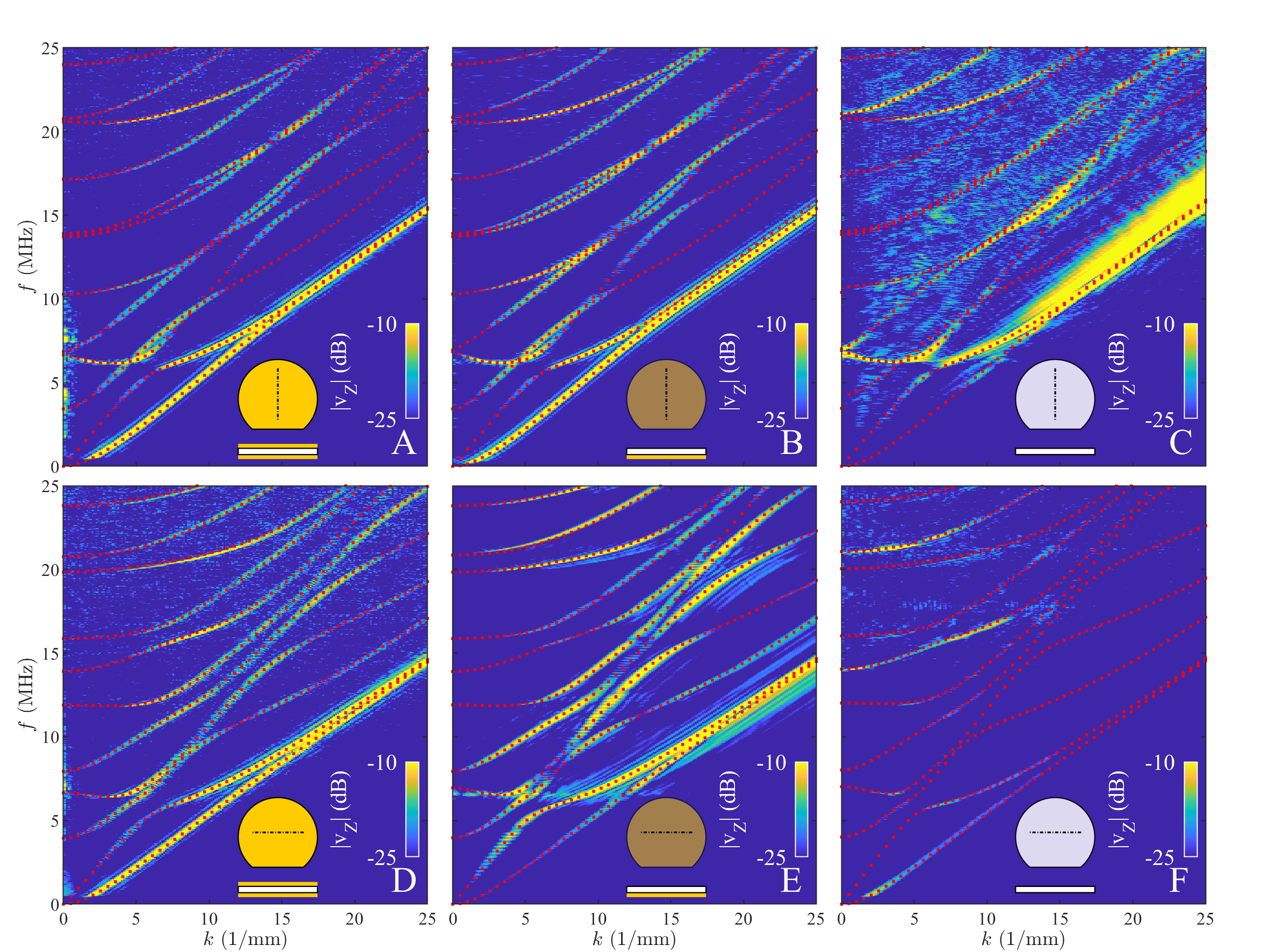}
    \caption{Comparison of the LUS measurements (intensity maps) and the SCM results (points). SH-like modes are omitted from the SCM results for clarity. The BC and propagation direction are indicated by the inset sketches. A:~shorted-shorted $X$-direction, B:~open-shorted $X$-direction, C:~open-open $X$-direction, D:~shorted-shorted $X + \qty{90}{\degree}$-direction, E:~open-shorted $X + \qty{90}{\degree}$-direction, F:~open-open $X + \qty{90}{\degree}$-direction.}
    \label{fig:LUS_SCM}
\end{figure*}
The experimental and theoretical results are superposed in \autoref{fig:LUS_SCM}.

The computations with the SCM are shown as red dots on top of the spectral magnitude obtained by the LUS experiments.
The insets indicate the direction of propagation and whether the surfaces are metallized or not.
Shear-horizontal waves, which are included in the SCM result (and shown faintly in \autoref{fig:dispersion_compare_bc}) are neither generated nor detected with the experimental setup and, thus, do not show. For clarity, these modes have been identified in the SCM solutions and omitted in the plot. 

Based on \autoref{fig:LUS_SCM}, excellent agreement is found between theory and experiment. Small deviations may be found in the highest order modes for the coated cases. The differences might be due to the finite thickness of the metallization, which is not represented in the SCM, or by an experimental deviation from the theoretical direction of propagation.
We stress, that the effect of different electrical BCs clearly shows in the experiment. For a qualitative assessment of the differences refer to Fig.~\ref{fig:dispersion_compare_bc}.

The results obtained on the uncoated plates (\autoref{fig:LUS_SCM}\,C and F) exhibit lower signal-to-noise ratio due to the workarounds described in \autoref{sec:experiments} but the acquired modes match well with the theory.

The configuration with the UV-laser results in an oval spot on the sample surface, rather than a line source. In this case (\autoref{fig:LUS_SCM}\,C), waves with arbitrary wave vector orientations are generated. All of them are potentially detected on the scan line, even when the wave vectors are not collinear to the scanning direction~\cite{kieferExtremeWaveSkewing2024}. While the main features of the measured dispersion shows good agreement with the calculations at fixed wave vector orientation, the blurred out regions in the experiment may be due to these wave skewing effects.

\section{\label{sec:conclusion}Conclusion}

We presented the SCM to calculate dispersion curves and mode shapes for guided electroelastic waves in piezoelectric plates. LUS experiments were conducted to measure dispersion curves on a \ce{LiNbO3} wafer. Comparison of theory and experiment for different directions and electrical BCs showed excellent agreement. From these results, material characterization based on an inverse problem seems feasible. 
A need for new characterization techniques is especially present in the case of deposited thin films and at high frequencies. The properties of thin films often differ from the bulk material, and depend on the growing process and layer thickness. 
We consider the possibility to measure a sample under different electrical BCs especially promising to disentangle elastic, piezoelectric and electric material properties. 

The SCM can be extended to multi-layered plates by the approach presented in Ref.~\cite{hernandoquintanillaModelingGuidedElastic2015}. This is particularly interesting for MEMS devices, which often consist of layered structures.

The model could then be used in combination with LUS measurements on different production steps of a stacked MEMS device, e.g., on the pure substrate of a SAW device and then on the deposited stack, including electrodes.
Previous work has proven that LUS can indeed be used to measure guided waves and resonances in the GHz range for material characterization purposes~\cite{Ryzy2023,Grunsteidl2020}. 
With the optical spot sizes that are achievable in practice, effective generation and detection of guided waves down to \SIrange{1}{2}{\micro\meter} wavelength is feasible.

As an alternative to SCM we discussed the SEM in Appendix~\ref{sec:spectral_elements}. It is based on the weak-form of the waveguide problem and is numerically superior to SCM but also somewhat more difficult to understand. An implementation of the SEM will be available in \texttt{GEWtool}~\cite{kieferGEWtool2023} for multi-layered piezoelectric plates.

\begin{acknowledgments}
This research was funded in whole or in part by the Austrian Science Fund FWF (P 33764). 
This project is co-financed by research subsidies granted by the government of Upper Austria (Wi-2021-303205/13-Au). 
Daniel A. Kiefer has received support under the program ``Investissements d'Avenir'' launched by the French Government under Reference No. ANR-10-LABX-24.
We thank Istvan Veres (Qorvo), Thomas Berer (Qorvo) and Jannis Bulling (BAM) for valueable discussions on the presented matter.
For the purpose of open access, the author has applied a CC BY public copyright license to any Author Accepted Manuscript version arising from this submission.
\end{acknowledgments}

\section*{Data Availability Statement}

The data that support the findings of this study are openly available in Zenodo at
\url{https://doi.org/10.5281/zenodo.13828765}, reference number [\cite{Zenodo}].

\appendix
\section{\label{sec:verification}Verifaction}
\begin{figure*}[tb]
    \centering
    \includegraphics[width=0.9\linewidth]{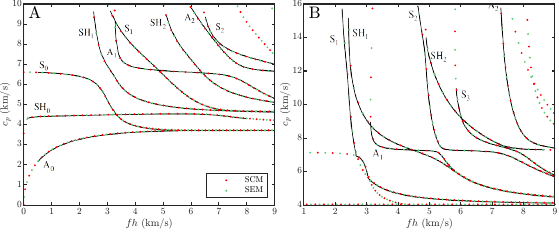}
    \caption{Verification of the SCM and the SEM with Kuznetsova et al.\cite{Kuznetsova2004} (black lines); A:~Y-cut, propagation in X-direction, B:~X-cut, propagation in Y+30°-direction}
    \label{fig:verKuz}
\end{figure*}

A comparison of phase velocity~$c_\mup{p} = \omega/k$ dispersion curves calculated with the SCM (and the SEM from Appendix~\ref{sec:spectral_elements}) to those published by Kuznetsova et al.~\cite{Kuznetsova2004} is shown in \autoref{fig:verKuz} and shows very good agreement. The latter authors employ root-finding of the characteristic equation. Two different crystal cuts of \ce{LiNbO3} and propagation directions are depicted.
Material constants are taken from \cite{kovacs1990} and rotated as needed before solving. Note that the root-searching method that serves as reference missed solutions, which is an inherent drawback of this method that is resolved by semi-analytical methods proposed here.

\begin{figure}[tb]
    \centering
    \includegraphics[width=7cm]{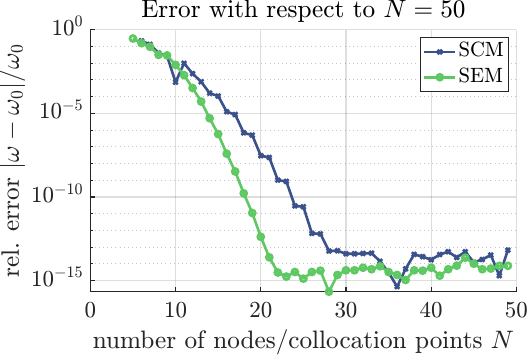}
    \caption{Convergence of the computed angular frequency~$\omega$ of the 12th mode at 20\,rad/mm ($\approx$ 21.7\,MHz) when increasing the number of collocation points or nodes~$N$.}
    \label{fig:convergence}
\end{figure}

The convergence of solutions is studied in Fig.~\ref{fig:convergence} for an arbitrarily selected mode. The wavenumber is fixed at 20\,rad/mm and we compute the angular frequency $\omega$ of the 12th mode (ordered in increasing $\omega$). As reference solution~$\omega_0$ we use our code with a high number of collocation points of $N = \text{50}$. As expected, the relative error $|\omega - \omega_0|/\omega_0$ is seen to decrease rapidly when increasing the number of collocation points~$N$. The corresponding convergence analysis with the SEM (Appendix~\ref{sec:spectral_elements}) is even more impressive.  For this rather high-order mode, 12 accurate digits are attained already with $N \ge 26$ for the SCM and $N \ge 20$ for the SEM.

\section{\label{sec:spectral_elements}Spectral element discretization}
The spectral element method (SEM) is numerically superior to SCM because it (i)~conserves the hermitian symmetry of the problem, (ii)~preserves the regularity of the operators and (iii)~converges faster. These properties also facilitate the postprocessing, e.g., computation of group velocities~\cite{finnvedenEvaluationModalDensity2004} and zero-group-velocity points~\cite{kieferComputingZerogroupvelocityPoints2023} as well as mode-sorting~\cite{gravenkampNotesOsculationsMode2023a}. SEM is akin to the finite element method but uses high-order polynomials defined on the entire domain as a basis, i.e., no meshing is necessary. In this section we sketch the SEM-discretization of the waveguide problem without providing theoretical details, for which we rather refer to the literature~\cite{hughesFiniteElementMethod2012,gravenkampNumericalApproachComputation2012}.

First, the boundary-value problem (\ref{eq:governingSystem}) and (\ref{eq:NeumannBCSystem}) needs to be cast into the corresponding weak form. To this end we introduce appropriate test functions~$\psi(z)$ corresponding to $\Psi(z)$, i.e., as a four-component vector function. 
Multiplying (\ref{eq:governingSystem}) with $\psi^T(z)$ from the left, integrating over the domain and performing a partial integration of the terms in $\mathcal{B}$ and $\mathcal{C}$ leads to the \emph{weak form of the waveguide problem}:

\begin{align}\label{eq:weakForm}
    \int_{-h/2}^{+h/2} \psi^T &\left[(\iu k)^2 \mathcal{A} + \iu k \mathcal{B}^T \partial_z + \omega^2 \mathcal{M} \right] \Psi \diff z \nonumber\\
    & - \int_{-h/2}^{+h/2} \partial_z \psi^T \left[\iu k \mathcal{B} + \mathcal{C} \partial_z \right] \Psi \diff z \nonumber\\
    & + \left. \left[ \psi^T (- \iu k  \mathcal{V}) \Psi\right]\right|_{-h/2}^{+h/2}
    = 0 \,. 
\end{align}
Thereby we have incorporated the Neumann BC from (\ref{eq:NeumannBCSystem}) into the bracketed boundary term.

Second, discretization is performed by approximating
\begin{align}\label{eq:approx}
    \Psi(z) &\approx \sum_j \underline{\Psi}_j P_j(z) \,,  &  
    \psi(z) &\approx \sum_i \underline{\psi}_i P_i(z) \,,
\end{align}
 as a weighted superposition of $N$ known polynomials $P_j(z)$ with $j \in \{1...N\}$ and 4-component vector coefficients $\underline{\Psi}_j$, $\underline{\psi}_i$. We choose Lagrange polynomials~$P_j(z)$ defined on the standard Gauß-Lobatto-Legendre points~\cite{hughesFiniteElementMethod2012}. 
 
 Accounting for (\ref{eq:approx}) in (\ref{eq:weakForm}), the equation must hold for arbitrary $\underline{\psi}_i$, so that we can write the discrete system

 \begin{align}\label{eq:SEMdiscrete}
    & [\, (\iu k)^2 \underbrace{ \mathcal{A} \otimes \mathrm{PP} }_{L_2} + \iu k \underbrace{ (\mathcal{B}^T \otimes \mathrm{PD} - \mathcal{B} \otimes \mathrm{PD}^T - \mathcal{V} \otimes \mathrm{BB}) }_{L_1} \nonumber\\
    & \quad \underbrace{ - \mathcal{C} \otimes \mathrm{DD} }_{L_0} + \omega^2 \underbrace{ \mathcal{M} \otimes \mathrm{PP} }_{M} \,] \Psi_{\mathup{d}}
    = 0 \,,
\end{align}
where $\Psi_{\mathup{d}} = [\underline{\Psi}_j]$, $j \in \{1...N\}$, and the $N \times N$-matrices
\begin{subequations}\label{eq:FEMmatrices}
\begin{align}
    \mathrm{PP}_{ij} &:= \int_{-h/2}^{+h/2} P_i P_j \diff z \,, \\
    \mathrm{PD}_{ij} &:= \int_{-h/2}^{+h/2} P_i \partial_z P_j \diff z \,, \\
    \mathrm{DD}_{ij} &:= \int_{-h/2}^{+h/2} \partial_z P_i \partial_z P_j \diff z \,, \\
    \mathrm{BB}_{ij} &:= \left. \left[ P_i P_j \right]\right|_{-h/2}^{+h/2} \,.
\end{align}
\end{subequations}
Note that in (\ref{eq:SEMdiscrete}) we have made use of the Kronecker product between $\mathrm{PP}$ and $\mathcal{A}$ (and similar for other terms), defined in block form as $\mathrm{PP} \otimes \mathcal{A} = [\mathrm{PP}_{ij} \, \mathcal{A}]$. $\mathcal{A} \otimes \mathrm{PP}$ is a permutation thereof that leads to an equally valid linear system. As usual, the matrices (\ref{eq:FEMmatrices}) are computed by numerical integration of the known basis polynomials. Dirichlet BCs, e.g., electrically shorted surfaces, can be implemented by removing the corresponding degree of freedom, as was done in Sec.~\ref{sec:scm}. The presented SEM procedure is implemented in \texttt{GEWtool}~\cite{kieferGEWtool2023}. A validation and convergence analysis can be found in Appendix~\ref{sec:verification}.

\section{\label{sec:material}Material parameters}
The material parameters used for the \ce{LiNbO3} of trigonal 3m symmetry are taken from Kovacs et al. \cite{kovacs1990}. We reproduce the values in Voigt-matrix notation below.

\noindent
Elastic stiffness under constant electric field:\\[5pt]
$ [\ten{c}] = \begin{bmatrix}
198.39&   54.72&   65.13&    7.88&       0  &         0 \\
      54.72&198.39& 65.13& -7.88&     0  &       0 \\
      65.13& 65.13&227.90&     0&     0  &       0 \\
       7.88& -7.88&     0  & 59.65&     0  &       0 \\
          0  &      0 &     0  &     0  & 59.65&  7.88 \\
          0  &      0 &     0  &     0  &  7.88& 71.835

\end{bmatrix}$\,GPa \\[5pt]
Piezoelectric coupling tensor (piezoelectric stresses):\\[5pt]
$ [\ten{e}] = \begin{bmatrix}
  0  &      0 &     0  &       0&    3.69&   -2.42\\
     -2.42&    2.42&     0  &    3.69&       0&       0\\
      0.30&    0.30&    1.77&       0&       0       0
\end{bmatrix}$\,C/m² \\[5pt]
Dielectric permittivity tensor under constant strain:\\[5pt]
$ [\ten{\epsilon}] = \epsilon_0 \begin{bmatrix}
45.6&      0&     0 \\
0&   45.6&     0 \\
0&      0&   26.3
\end{bmatrix}, \quad \epsilon_0 \approx 8.854188\,\text{F/m} $ \\[5pt]
Mass density: $\rho = $ 4628\,kg/m³

\bibliography{bibliography,bibliographyDaniel}

\end{document}